\documentclass[12pt]{article}
\pdfoutput=1

\usepackage{draft,hyperref,tikz,cancel,subfig,float}
\usepackage{multirow}
\usepackage{soul}
\usepackage[nosort]{cite}
\usepackage{amsmath,mathtools}
\usepackage{amsfonts}
\usepackage{amsthm}
\usepackage{slashed}
\usepackage{xcolor}
    \usepackage{graphicx}

\usetikzlibrary{snakes}
\usetikzlibrary{shapes.misc}
    \newcommand{\secref}[1]{\S\ref{#1}}
    \newcommand{\figref}[1]{Figure~\ref{#1}}

    \def\ie{\begin{equation}\begin{aligned}}
    \def\fe{\end{aligned}\end{equation}}
     
    \def\ria{\rightarrow}
    \newcommand{\E}{{\epsilon}}
    
    \newcommand{\cH}{{\mathcal H}}
    
    \newcommand{\cI}{{\mathcal I}}

\def\xTB{$\times$} 

\begin{document}

\begin{titlepage}

\begin{center}

\hfill \\
\hfill \\
\vskip 1cm

\title{OPE of line defects in 5d $E_n$ SCFT}

\author{Jihwan Oh}
\address{Mathematical Institute, University of Oxford,
Woodstock Road, Oxford, OX2 6GG, United Kingdom}

\end{center}

\abstract{5d ray index is the 5d superconformal index in the presence of a ray-like defect. An elementary ray defect flows to a fundamental Wilson ray in IR gauge theory. In 5d $\cN=1$ superconformal field theory(SCFT) with $E_n$ global symmetry, we compute the ray index associated with an adjoint Wilson ray.  We show that the operators that appear in the index are non-trivially charged under the center of the global symmetry $E_n$. The charge under the center being twice compared with the center charge carried by the elementary ray operator indicates a non-trivial OPE between the ray-like defects in the UV SCFT.} 
\end{titlepage}

\tableofcontents
\section{Introduction and summary}
One of the most promising ways to study a strongly coupled supersymmetric quantum field theory is supersymmetric localization \cite{Pestun:2007rz}. Utilizing this technique, one can compute the superconformal index \cite{Kinney:2005ej,Bhattacharya:2008zy,Kim:2012gu}, defined as $S^{d-1}
\times S^1$ partition function of d-dimensional superconformal field theory. The index counts a certain set of local BPS operators. Once we introduce a non-local defect in the theory, such as a ray operator $\cL_i$ with an open end, the previous Hilbert space defined without the ray operator is deformed. The authors of \cite{Chang:2016iji} defined and computed the relevant index that counts states in the deformed Hilbert space in the context of 5d $\cN=1$ superconformal field theory (SCFT) with $E_n$ global symmetry; the index was named ray operator index. The goal of this work is to provide an evidence of the existence of the line defect OPE $\cL_1\times
\cL_2\ria\cL_3$ by comparing ray operator index associated to each ray operator $\cL_i$. 

The 5d SCFT is believed to be a UV fixed point of IR 5d $\cN=1$ $G=Sp(1)$ SYM with fundamental matters \cite{Seiberg:1996bd} and a superconformal line defect is UV ancestor of a supersymmetric Wilson line in the IR gauge theory. One way to classify the superconformal line defect is to use the 5d superconformal algebra \cite{Agmon:2020pde}. This is natural in a sense that it only uses the intrinsic information of the SCFT itself. However, since we rely on the localization technique that uses the IR gauge theory, to make contact with the computation, it looks more useful to label the defect by property of its IR counterpart: the representation of the IR Wilson ray under the IR gauge symmetry\footnote{Note, however, that the gauge symmetry does not exist at the UV conformal fixed point.}. Let us then denote superconformal line defect with its IR descendant labeled by a representation $R$ as $\cL_{R}$ and the relevant ray operator index as $\la\cL_R\ra$. 

For the IR gauge theory with $G=Sp(1)$, the simplest nontrivial Wilson ray operator is labeled by the fundamental representation. Hence, we can think $\cL_{fund}$ as the simplest nontrivial superconformal line operator. An important property of $\cL_{fund}$ is that the local operator that ends on $\cL_{fund}$ is charged non-trivially under the center of the global symmetry $E_n$ of the SCFT. Let us denote the center charge $c$. This can be explicitly observed and computed by analyzing the index $\la\cL_{fund}\ra$ \cite{Chang:2016iji}.

We will compute $\la\cL_{adjoint}\ra$ and show that $\cL_{adjoint}$ carries $2c$ by enumerating all global symmetry representations that appear in $\la\cL_{adjoint}\ra$. This result strongly indicates the existence of a non-trivial OPE between two superconformal line operators $\cL_{fund}$ that leads to $\cL_{adjoint}$:
\ie
\cL_{fund}\times\cL_{fund}\ria\cL_{adjoint}.
\fe

In \secref{review}, we will review definitions and a brief computation procedure of 5d superconformal index and ray operator index. In \secref{result}, we will present the adjoint ray operator index and show that the relevant local BPS operators, where the ray ends, carry double amount of charges under the $\bZ_{9-n}$ center of the $E_n$ global symmetry compared to the center charge of the local operators at the end of the fundamental ray operator.
\section{Ray Operator Index and Superconformal Index}\label{review}
\begin{figure}[H]
\centering
\includegraphics[width=11cm]{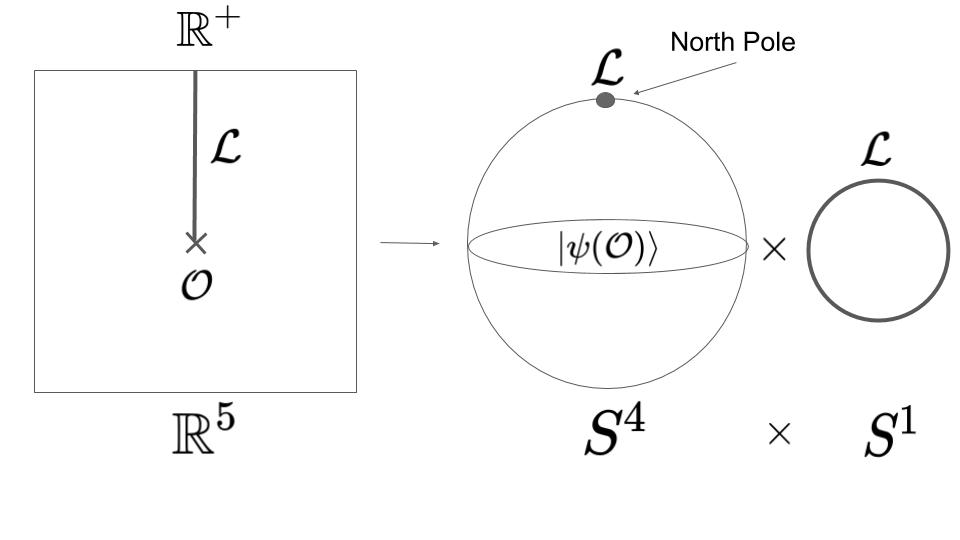}
\caption{Left: Theory on $\mathbb{R}^5$ with a line defect $\mathcal{L}$ stretched along $\mathbb{R}^+$ ending at the origin of $\mathbb{R}^5$ with a local junction operator $\mathcal{O}$ sitting at the end point. Right: Theory on $S^4\times S^1$ with $\mathcal{L}$ wrapping $S^1$ and sitting at the north pole of $S^4$.
The system on the left is related to the system on the right by a conformal transformation, with the local junction operator $\mathcal{O}$ being mapped to a state $|\psi(\mathcal{O})\rangle$ on $S^4$ (with $\mathcal{L}$ sitting at the north pole).}
\label{rayops}
 \centering
\end{figure}

Ray operator index is Witten index\cite{Witten:1982df} that counts states in the Hilbert space on $S^4$ of $5d$ SCFTs in the presence of a line defect $\mathcal{L}$ piercing the spatial $S^4$ at a point. Due to state-operator correspondence, this is the same as counting local junction operators $\mathcal{O}$ in the $5d$ SCFT on $\mathbb{R}^5$ where the line defect $\mathcal{L}$ stretched along $\mathbb{R}^+$ can end -- this is depicted in figure \ref{rayops}.

As $5d$ SCFTs do not admit a Lagrangian description, we need to rely on the RG flow invariance of BPS protected operators and compute the index using localization techniques \cite{Pestun:2007rz,Kim:2012gu}. These are applied to the IR $5d$ $\cN=1$ gauge theory description of the UV SCFTs. The line defect of the UV SCFT reduces to a Wilson line defect in the IR non-abelian gauge theory. In this section, we sketch the procedure to compute the path integral of the IR gauge theory on $S^4\times S^1$ in the presence of a Wilson line defect wrapped along $\{ \text{North Pole} \}\times S^1\subset S^4\times S^1$.
\noindent
Before computing the index using  localization techniques, it is helpful to recall the Hamiltonian definition of the index to set up notation. The ray operator index can be viewed as a character of the line defect enriched Hilbert space $\mathcal{H}_{\text{ray}}$ (viewed as a module over the superconformal algebra $F(4)$ whose Cartan sub-algebra is generated by $\Delta, J_\pm, J_R$, which are generators of dilatation, two rotations, and $SU(2)_R$ symmetry):
\ie
&\mathcal{I}^{\text{ray}}(\E_+,\E_-,m_i,M)=Tr_{\mathcal{H}_{\text{ray}}}\left[(-1)^Fe^{-\B\Delta}e^{-2\E_+(J_++J_R)-2\E_-J_-}e^{-\sum F_im_i}e^{\Pi M}\right],
\fe
where $F$ is the fermion number. $\E_\pm$, $m_i$, $M$ are chemical potentials for $J_\pm$, flavor symmetry generators $F_i$, a certain $Sp(1)'$ symmetry generator\footnote{We will be more explicit about $\Pi$ in \eqref{SQMWittenIndex}.} $\Pi$ relevant to an implementation of the ray operator, and $\B$ is the radius of the time circle $S^1$, where the trace is taken over $\mathcal{H}_{\text{ray}}$. Due to state-operator correspondence in conformal field theory, the Hilbert space is isomorphic to the space of local operators on a flat spacetime sitting at an end of the line defect. 

The practical computation of the above index defined in the Hamiltonian formalism is through a Lagrangian way -- we compute it using the path integral of the IR gauge theory, whose UV theory is the SCFT. The conjectured RG-flow invariance of the content of the BPS operators\footnote{In particular, we are counting 1/8-BPS operators annihilated by $Q=Q^1_2$ and $S=S^2_1$, where $Q^A_m$, $S^n_B$ are (conformal)supercharges with $A,B$ being $SU(2)_R$ spinor index and $m,n$ $SO(5)$ vector indices. These operators satisfy the following unitarity bound $\{Q,S\}=\Delta-2J_+-3J_R\geq0$.} ensures that the result obtained in this way gives the index of the UV SCFT. 

Concretely, we consider a $5d$ $\mathcal{N}=1$ supersymmetric gauge theory with a gauge group $G=Sp(N)$, a vector multiplet $\cV$, $N_f$ fundamental hypermultiplets $\cH_i$, and one anti-symmetric hypermultiplet $\mathcal{H}_{AS}$. Given the field content, we can utilize \cite{Pestun:2007rz,Kim:2012gu} to compute the path integral using supersymmetric localization. More explicitly, we deform the $5d$ gauge theory Lagrangian by some Q-exact term, $\mathcal{L}'=\mathcal{L}+t\{Q,V\}$, where $t$ is some constant and $V$ is a function of fields in the supermultiplets $\cV$, $\cH_i$, $\mathcal{H}_{AS}$. Because the theory is supersymmetric, the path integral does not change after the deformation with any $t$. By taking $t\rightarrow\infty$, the path integral domain localizes to the moduli space of solutions of the saddle point equation $\{Q,V\}=0$. In the particular geometry $S^4\times S^1$ that we are studying, the localization locus turns out to be (1) instanton$(F=\star F)$, (2) anti-instanton$(F=-\star F)$ configuration on North and South pole of $S^4$, which are $SO(4)$ fixed points and (3) perturbative modes of the Lagrangian fields on the rest of $S^4$. We can summarize the localization result as follows:
\ie\label{inte}
\mathcal{I}=\int[d\alpha]Z_{\text{pert}}Z_{\text{inst}}Z_{\text{anti-inst}}\,,
\fe 
where $[d\alpha]$ is $Sp(N)$ Haar measure, given by
\ie\label{spnhaar}
\left[d\alpha\right]=&\frac{2^N}{N!}\left[\prod^N_{i=1}\frac{d\alpha_i}{2\pi}\sin^2\A_i\right]\prod_{i<j}\left[2\sin\left(\frac{\A_i-\A_j}{2}\right)\right]^2\left[2\sin\left(\frac{\A_i+\A_j}{2}\right)\right]^2 \,.
\fe
We will explain each factor in \eqref{inte} soon. Note that due to the presence of the line defect at the North Pole, the moduli space of instanton will be deformed, so we will need to take extra care later when we embed the setting into string theory to compute $Z_{\text{inst}}$. On the other hand, the South Pole contribution is purely from un-deformed anti-instanton moduli space.

$Z_{\text{pert}}$ essentially encodes all the $5d$ operators that can be built out of elementary Lagrangian fields that are in the Q-cohomology. If we denote the single-letter index as
\ie
\mathfrak{f}=\sum_{\text{letters}}(-1)^Ft^{2(J_++J_R)}u^{2J_-}v^{f}m_i^{f_i}=\mathfrak{f}_{\text{vec}}+\mathfrak{f}_{\text{asym}}+\mathfrak{f}_{\text{fund}}
\fe
where $t=e^{-\E_+}$, $u=e^{-\E_-}$, $v=e^{-m}$, $m_i=e^{-F_i}$. $f$, $f_i$ are the generators of the flavor symmetry associated with the anti-symmetric and fundamental hypermultiplets, respectively, and 
\ie
\mathfrak{f}_{\text{vec}}&=-\frac{t(u+u^{-1})}{(1-tu)(1-tu^{-1})},\quad \mathfrak{f}_{\text{asym}}=\frac{t}{(1-tu)(1-tu^{-1})}(v+v^{-1}),\\
\mathfrak{f}_{\text{fund}}&=\frac{t}{(1-tu)(1-tu^{-1})}\sum^{N_f}_{i=1}2\cosh m_i,
\fe
where $N_f$ is the number of fundamental hypermultiplets.

We may write the multiletter index as a Plethystic exponential, and this is equivalent to the 1-loop determinants of $5d$ perturbative modes, $Z_{\text{pert}}$.
\ie
Z_{\text{pert}}=P.E[\mathfrak{f}]=\exp\left[\sum_{\bf{R}}\sum^\infty_{n=1}\mathfrak{f}_{\bf{R}}(t^n,u^n,v^n)\chi_{\bf{R}}(w_i^n)\right],
\fe
where
\ie
\chi_{\text{vec}}(w_i)=&\bigg[\sum^N_{i<j}\left(\frac{1}{w_iw_j}+\frac{w_j}{w_i}+\frac{w_i}{w_j}+w_iw_j\right)+\sum^N_{i=1}\left(w_i^{-2}+w_i^2\right)+N \bigg]\\
\chi_{\text{asym}}(w_i)=&\left[\sum^N_{i<j}\left(\frac{1}{w_iw_j}+\frac{w_j}{w_i}+\frac{w_i}{w_j}+w_iw_j\right)+N\right] \,.
\fe
Note that $w_i$ and $\A_i$ appearing in \eqref{spnhaar} are related as $w_i=e^{i\A_i}$.

The set of BPS instanton operators that are captured by $Z_{\text{inst}}$ and $Z_{\text{anti-inst}}$ are graded by the instanton number $k$. $Z_{\text{inst}}$, also known as Nekrasov partition function \cite{Nekrasov:2002qd}, can be expressed as a generating series in $q$, where each term is relevant for the subsector of the set of BPS operators with charge $k$ under the $U(1)$ topological symmetry of 5d gauge theory:
\ie
Z_{\text{inst}}=1+\sum_{K=1}^\infty q^KZ^K_{\text{inst}} \,.
\fe
Each of $Z^K_{\text{inst}}$ needs to be computed separately.

\begin{figure}[H]
\centering
  \vspace{-20pt}
\includegraphics[width=11cm]{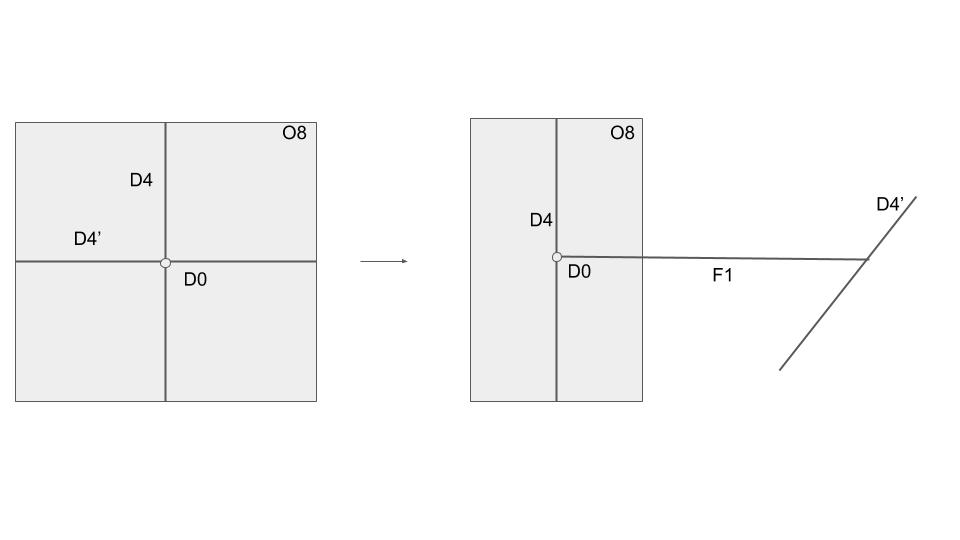}
  \vspace{-35pt}
\caption{Before expanding in $x$ at $\infty$, the index captures BPS operators associated with the left figure. After the expansion, the index counts BPS operators associated with the right figure.}
\label{massivef1}
 \centering
\end{figure}

To compute the $K$-th instanton partition function $Z^K_{\text{inst}}$, it is convenient to embed the system into string theory, where the $5d$ gauge theory with gauge group $Sp(N)$ and its $K$ instantons are realized as $N$ D4-branes and 
$K$ D0-branes on an O8-plane. Given this embedding, we also need to explain the string theory realization of the line defect. It was noticed in \cite{Chang:2016iji,Tong:2014cha,Kim:2016qqs} that the ray-like defect can be implemented as a trajectory of one end of the fundamental string between the D4-branes and an additional D4$'$-brane, whose directions in the spacetime can be found in the following table.
\begin{table}[H]
\centering
\begin{tabular}{|c|cccccccccc|l|} \hline
   &  0 &  1 &  2 &  3 &  4 &  5 &  6 &  7 &  8 &  9  \\
\hline\hline
O8 &\xTB&\xTB&\xTB&\xTB&\xTB&\xTB&\xTB&\xTB&\xTB&   \\
\hline
D4 &\xTB&\xTB&\xTB&\xTB&\xTB&    &    &    &    &     \\
\hline
F1 &\xTB&   &   &   &   &    &    &    &    &\xTB\\
\hline
D$4'$ &\xTB&    &    &    &    &\xTB&\xTB&\xTB&\xTB&   \\
\hline
\end{tabular}
\caption{The directions of the various branes.}
\label{table:BraneDirections}
\end{table}
\noindent For better graphical illustration, see \figref{massivef1}. We will come back to a more familiar IR gauge theory description of the line defect as a Wilson ray in the next section.

$\tilde Z^k_{\text{inst}}$\footnote{We will explain the difference between $\tilde Z^k_{inst}$ and $Z^k_{inst}$ shortly.} is then a partition function of  the $K$ D0-brane worldvolume theory which is 1d $\cN=4$, $G=O(K)$, $G_F=Sp(N)$ supersymmetric quantum mechanics. The field content is determined by the proper quantization of all possible strings that connects different D-branes in the system. 
 
 Formally, the D0-brane partition function can be written as a form of Witten index as
\ie\label{SQMWittenIndex}
\tilde Z^k_{\text{inst}}=Tr_{\mathcal{H}_{\text{QM}}}\left[(-1)^Fe^{\beta\{Q,Q^\dagger\}}t^{2(J_++J_R)}u^{2J_-}v^{2J'_R}w^{2\Pi_i}x^{\Pi}\right]. 
\fe
where the newly introduced generators(or at the same time charges under those) $J'_R$, $\Pi_i$, and $\Pi$ are the Cartan generators of $SU(2)_{R'}$\footnote{$\cN=4$ SQM has two $SU(2)$ R-symmetries; one is $SU(2)_R$ and the other is $SU(2)_{R}'$} that rotates anti-symmetric hypermultiplet in 5d, $Sp(N)$ of $N$ D4-branes, $Sp(1)'$ of D4$'$-brane, respectively, and $\mathcal{H}_{\text{QM}}$ is the Hilbert space of the quantum mechanics. 

The Witten index can again be computed by the supersymmetric localization of the quantum mechanical path integral. One should remember that since the gauge group of the quantum mechanics $G=O(K)$ has two disjoint components, and we need to compute SQM partition function separately for each of $O(K)_\pm$. As the procedure and the result were discussed in the literature \cite{Hwang:2014uwa,Hori:2014tda} extensively, we will just write down the final expression and use it in the computation in the next section. 

To sum up, we may write  $\tilde{Z}^k_{\text{inst}}$ as follows. 
 \ie\label{kinstexplicit}
 \tilde{Z}^k_{\text{inst}}(t,u,v,w_i,x)=&\frac{1}{2}\left(\tilde{Z}^k_+(t,u,v,w_i,x)+\tilde{Z}^k_-(t,u,v,w_i,x)\right)\\
 \tilde{Z}^k_\pm(t,u,v,w_i,x)=&\frac{1}{\rvert W\rvert}\oint\prod_{i=1}^{[K/2]}[d\phi_i]Z^{\pm,k}_{\text{D0-D0}}Z^{\pm,k}_{\text{D0-D4}}Z^{\pm,k}_{\text{D0-D4$'$}}Z_{\text{D4-D4$'$}}
 \fe
where $k=2n+\chi$ with $\chi=0,1$ and
\ie
\rvert W\rvert^{\chi=0}_+=\frac{1}{2^{n-1}n!},~\rvert W\rvert^{\chi=1}_+=\frac{1}{2^nn!},~\rvert W\rvert^{\chi=0}_-=\frac{1}{2^{n-1}(n-1)!},~\rvert W\rvert^{\chi=1}_-=\frac{1}{2^nn!}.
\fe
The expression in the square bracket is a Haar measure of $O(K)$\footnote{It can be found in Appendix E of \cite{Kim:2012gu}.}, and the new fugacity $x=e^{-iM}$ is that of D4$'$-brane. $Z^{\pm,k}_{\text{Di-Dj}}$ stand for 1-loop determinants of quantum mechanical fields that are from quantization of $Di-Dj$ strings in $K$ D0-branes sector. The various 1-loop determinants were computed in \cite{Kim:2012gu,Chang:2016iji}, and they take the following form:
 \ie Z_{\text{Da-Db}}=\frac{\prod_{j=1}^{n_1}\sinh(\vec\rho_j\cdot\vec\phi+f_j(\E_+,\E_-,\A_i,m,M))}{\prod_{k=1}^{n_2}\sinh(\vec\rho_k\cdot\vec\phi+f_k(\E_+,\E_-,\A_i,m,M))},
 \fe
where $\vec\rho_j$, $\vec\rho_k$ are in the root lattice of $G=O(K)$, $\vec\phi=(\phi_1,\ldots,\phi_{[K/2]})$, and $n_i\in\bZ^+$. We direct the reader to \cite{Kim:2012gu,Chang:2016iji,Hwang:2014uwa,Hori:2014tda} for the explicit expressions that we used in our calculation. Looking at the expression, we notice there are many poles in the integrand. We used a contour prescription called Jeffrey-Kirwan(JK) residue formula \cite{Benini:2013nda,Benini:2013xpa} to evaluate the $d\phi_i$ integral. 

Given all $\tilde{Z}^k_{\text{inst}}$, let us define $\tilde{Z}_{\text{inst}}$ as a generating series
\ie
\tilde{Z}_{\text{inst}}(q,t,u,v,w_i,x)=\sum_{k=0}q^k\tilde{Z}^k_{\text{inst}}(t,u,v,w_i,x).
\fe
Here $\tilde{Z}^0_{\text{inst}}\equiv Z_{\text{D4-D4$'$}}$, since this factor is not a function of $\phi_i$, so it can be factored out of each $\phi_i$-integral \eqref{kinstexplicit} and it becomes an overall factor of the entire series.

Although the string theory embedding has the advantage of translating the problem of computing the instanton partition function to the D0-brane quantum mechanics partition function, it creates a subtlety involving the ``extra" states, which we need to decouple from $\tilde{Z}_{\text{inst}}$ to get the true instanton partition function $Z_{\text{inst}}$, as pointed out in \cite{Hwang:2014uwa}. The source of extra states is the D0-branes that are unbound to D4-branes. This definition simply leads to the following expression for the extra partition function $Z_{\text{extra}}^k$ whose integrand does not involve any 1-loop determinants of the fields obtained from D4-related strings:
 \ie\label{extra}
{Z}^k_{\text{extra}}(t,u,v,x)&=\frac{1}{2}({Z}^{k,+}_{\text{extra}}(t,u,v,x)+{Z}^{k,-}_{\text{extra}}(t,u,v,x))\\
{Z}^{k,\pm}_{\text{extra}}(t,u,v,x)&=\frac{1}{\rvert W\rvert}\oint\prod_{i=1}^{[K/2]}[d\phi_i]Z^{\pm,k}_{\text{D0-D0}}Z^{\pm,k}_{\text{D0-D4$'$}} \,.
 \fe
Let us also define $Z_{\text{extra}}$ as a generating series
\ie
Z_{\text{extra}}(q,t,u,v,x)=1+\sum_{k=1}q^kZ^k_{\text{extra}}(t,u,v,x) \,.
\fe
The true instanton partition function $Z_{\text{inst}}$, which is used in the 5d index computation, is then obtained by eliminating the extra states from $\tilde{Z}_{\text{inst}}$
\ie
Z_{\text{inst}}(q,t,u,v,w_i,x)=\frac{\tilde{Z}_{\text{inst}}(q,t,u,v,w_i,x)}{Z_{\text{extra}}(q,t,u,v,x)}\,.
\fe
As a side remark, it was first noticed in \cite{Kim:2016qqs}, this function is related to the Nekrasov's qq character \cite{Nekrasov:2015wsu}. And for the case of $G=U(K)$, $G_F=U(N)$(flavor symmetry group), and a general number of D4$'$-branes, the function has a compact analytic form, written down in \cite{Agarwal:2018tso}.

Now, let us turn to $Z_{\text{anti-inst}}$. In contrast to $Z_{\text{inst}}$, which is modified by the ray-like defect or D4$'$-brane in the string theory picture, $Z_{\text{anti-inst}}$ is associated to the pure instanton moduli space. In the D0-brane QM language, we simply exclude D4$'$-brane in the above set-up and follow the same procedure by simultaneously taking inverse of $q$. Of course, here we also decouple extra unbound D0-branes to D4-branes
\ie
Z_{\text{anti-inst}}&(t,u,v,q,w_i)=\frac{1+\sum_{k=1}q^{-k}\tilde{Z}^k_{\text{anti-inst}}(t,u,v,w_i)}{1+\sum_{k=1}q^{-k}\tilde{Z}^k_{\text{extra}}(t,u,v,w_i)}\\
\tilde{Z}^k_{\text{anti-inst}}&=\frac{1}{2}({\tilde Z}^{k,+}_{\text{anti-inst}}+{\tilde Z}^{k,-}_{\text{anti-inst}}),\quad\tilde{Z}^k_{\text{extra}}=\frac{1}{2}({\tilde Z}^{k,+}_{\text{extra}}+{\tilde Z}^{k,-}_{\text{extra}})\\
{\tilde Z}^{k,\pm}_{\text{anti-inst}}&=\frac{1}{\rvert W\rvert}\oint\prod_{i=1}^{[K/2]}[d\phi_i]Z^{\pm,k}_{\text{D0-D0}}Z^{\pm,k}_{\text{D0-D4}},\quad\tilde{Z}^{k,\pm}_{\text{extra}}=\frac{1}{\rvert W\rvert}\oint\prod_{i=1}^{[K/2]}[d\phi_i]Z^{\pm,k}_{\text{D0-D0}} \,.
\fe

Now that we have all ingredients for the final integral, so let us collect the factors and rewrite the integral
\ie\label{finalinte}
\mathcal{I}^{\text{ray}}(t,u,v,q,x)=\int[dw_i]Z_{\text{pert}}Z_{\text{inst}}Z_{\text{anti-inst}}
\fe
It is actually instructive to factor out $Z_{\text{D4-D4$'$}}(w_i,x)$ from $Z_{\text{inst}}$ to analyze the behavior of the ray index in the first few leading powers of $x$. Let us denote $\bar{Z}_{\text{inst}}=Z_{\text{\text{inst}}}/Z_{\text{D4-D4$'$}}$.
\ie\label{d4d4pexp}
\mathcal{I}^{\text{ray}}(t,u,v,q,x)=&\int[dw_i]Z_{\text{D4-D4$'$}}(w_i,x)Z_{\text{pert}}\bar{Z}_{\text{inst}}Z_{\text{anti-inst}}\\
=&\int[dw_i]\prod^N_{j=1}\left(x^{-1}-w_j-w_j^{-1}+x\right)Z_{\text{pert}}\bar{Z}_{\text{inst}}Z_{\text{anti-inst}} \,.
\fe
From the last line, we notice the lowest order in $x$ is $x^{-N}$. It was observed in \cite{Chang:2016iji} that for $N=1$, the lowest order term in $x$, $O(x^{-1})$ reproduces the superconformal index, and was argued further that the next order in $x$, $O(x^{-1+1})$ is the fundamental ray operator index. In other words,
\ie\label{finalexp}
\mathcal{I}^{\text{ray}}=x^{-1}\big(\mathcal{I}_{\text{SCI}}+x\mathcal{I}^{\bf{fund}}_{\text{ray}}+\ldots\big) \,.
\fe
Note that we expand $\mathcal{I}^{\text{ray}}$ in $x=e^M$ at $x=\infty$ consistent with the definition of the ray-like defect created by the very heavy fundamental strings stretched between the D4-branes and a D4$'$-brane.
\noindent
\section{Adjoint ray as OPE of fundamental ray}\label{result}
\subsection{Defect Hilbert Space and OPE of Ray Operators}
Recall that the open Wilson ray is in the IR gauge theory needs to end with a local operator to preserve the gauge invariance.
\ie
\mathcal{L}_{\mathcal{O}_{\mathfrak{R}}}=P\exp\left[i\int_0^\infty\left(A_0+\Phi\right)dx^0\right]{\mathcal{O}_{\mathfrak{R}}}(0) \,,
\fe
where $\mathcal{O}_{\mathfrak{R}}(0)$ denotes a local operator at the origin on which the open Wilson ray labeled by the representation $\mathfrak{R}$ may end so that the combination is invariant under the IR gauge symmetry. The ray index counts the UV image of $\cL_{\cO_{\mathfrak{R}}}$, or simply that of $\cO_{\mathfrak{R}}(0)$. In this subsection, we will illustrate how to extract various $\la\cL_{\cO_{\mathfrak{R}}}\ra$ from $\cI^{\text{ray}}$.

To track the UV image of $\mathcal{L}_{\mathcal{O}_{\mathfrak{R}}}$ in \eqref{finalexp}, we need to track the gauge fugacity $w_i$ (of the D4-brane), or more precisely $\chi_{\mathfrak{R}}(w_i)$. However, since we have already integrated out $w_i$'s in the previous step \eqref{d4d4pexp}, the final expression $\cI^{\text{ray}}$ \eqref{finalexp}  does not depend on the gauge fugacity $w_i$. Rather, we may equivalently track the surviving fugacity $x$(of the D4$'$-brane). We can validate this alternative tracking by recalling the UV definition of the Wilson ray. It is given by the trajectory of the end of the D4-D4$'$ string in the D4-brane worldvolume theory, which is the $5d$ UV SCFT. The other end of the same string is labeled by $x$, and the D4-D4$'$ string(or the field obtained from quantizing the string) transforms as a bi-fundamental representation under $Sp(N)_{\text{D4}}\times Sp(1)_{\text{D4$'$}}$, which endows equal status for $w$ and $x$. 

It is important to notice the final index formula only contains the BPS operators that transform as antisymmetric representations under $Sp(N)_{\text{D4}}\times Sp(1)_{\text{D4$'$}}$, since D4-D4$'$ strings are fermionic. For instance, for $N=1$, $O(x^{-N+2})$ terms encode ray operators that transform under $Sp(1)_{D4}\times Sp(1)_{D4'}$ as
\ie
(\bf{2},\bf{2})\wedge(\bf{2},\bf{2})=(\bf{3},\bf{1})\oplus(\bf{1},\bf{3}).
\fe
However, since those operators transformed as ${(\bf{1},\bf{3})}$ are essentially from strings that have both ends on the D4$'$-brane, it has already been eliminated when we decoupled $Z_{\text{extra}}$. Hence, we are left with ${(\bf{3},\bf{1})}={({\bf{Adj},1})}$, which we call the adjoint ray operator. In other words,
\ie\label{finalindex}
\mathcal{I}^{\text{ray}}_{Sp(1)}=x^{-N}\big(\mathcal{I}_{\text{SCI}}+x\mathcal{I}^{\bf{fund}}_{\text{ray}}+x^{2}\mathcal{I}^{\bf{adj}}_{\text{ray}}+\ldots\big) \,.
\fe
In our convention, this means that
\ie
\cI_{\text{SCI}}=\la\cL_{\cO_{\bf{1}}}\ra,\quad\cI^{\bf{fund}}_{\text{ray}}=\la\cL_{\cO_{\bf{fund}}}\ra,\quad\cI^{\bf{adj}}_{\text{ray}}=\la\cL_{\cO_{\bf{adj}}}\ra.
\fe
Similarly, we can expect that more complicated representations will appear as coefficients of $x^n$(with $n\geq2$) in the parenthesis\footnote{It will be nice to use the technique discussed in \cite{Gaiotto:2015una} and do the computation for ray operator index for different $\mathfrak{R}$.}.

Let us analyze the underlying physical meaning of the above argument and discuss OPE of $\cL_{\cO_{\bf{fund}}}$. The generating series \eqref{finalexp}, which we have called as $\cI^{\text{ray}}$, encodes the full information of the ray operator index at the UV fixed point. Each coefficient $a_k$ of $x^k$ carries a partial information of the full ray operator index, and we can label each of them with its IR image\footnote{Similar idea was dicsussed in \cite{Gaiotto:2010be,Cordova:2016uwk,Neitzke:2017cxz}.}: k-th tensor product of the fundamental Wilson rays of the IR gauge theory. In other words, the defect Hilbert space at the UV fixed point is a graded Hilbert space with its grading given by the number $k$:
\ie\label{decomp}
\mathcal{H}_{defect}=\bigoplus_{k=1}\mathcal{H}_{defect}^k.
\fe

The states in each $\mathcal{H}_{defect}^k$ correspond to the BPS operators due to the state/operator correspondence, and the set of BPS operators has a structure of a ring with a multiplication defined by a usual operator product. A natural question is if the grading is preserved under the multiplication of the BPS operators in different sectors $\mathcal{H}_{defect}^{k_1}$, $\mathcal{H}_{defect}^{k_2}$ with $k_1\neq k_2$. A possible starting point of the analysis is the product between states/operators in $k=1$ sector(fundamental ray index), and see if it yields those in $k=2$ sector(adjoint ray index).

The strategy of testing the proposal should be to compare the $E_{N_f+1}$ representations that appear in a product of operators in the $k=1$ sector and that appear in the $k=2$ sector. This is in principle a possible task; we indeed obtain all $E_{N_f+1}$ representations of the $k=2$ sector. However, for the sake of the comparison, we rather take a detour, which reduces a big amount of work. 

The representative property of the ray operator index is that $E_{N_f+1}$ representations that appear in the ray index are charged under the center $Z_{8-N_f}$ of the global symmetry $E_{N_f+1}$ \cite{Chang:2016iji}. We will simply test if the $Z_{8-N_f}$ center charges of representations that appear in $k=2$ sector matches with two times that of $k=1$ sector. Our first example$(N_f=0)$ does not have a quark, where the fundamental Wilson ray can end. Hence, the $k=1$ sector is absent. However, in the following examples we include fundamental hypermultiplets, which provide quarks and we can fill in the $k=1$ sector and test the above proposal. 

This method suffices for our purpose to propose that the ray number grading $k$ preserves under the OPE of ray operators in different sectors. In the following subsection, we will do the following.
\begin{itemize}
    \item We will present adjoint ray operator index for $N_f=0$ result for $G=Sp(1)$, $Sp(2)$, $Sp(3)$. Note that we have included up to 3 instantons in the final expression of $Z_{\text{inst}}$ and $Z_{\text{anti-inst}}$ in the integrand \eqref{finalinte}. Since we are only interested in low spin operators, i.e. low power of t terms, the higher instantons are irrelevant, as they only contribute to the higher spin operators. This fact guarantees that including up to 3 instantons yields the full answer, if we restrict our attention to low powers of $t$.
    \item Test our OPE proposal for $G=Sp(1)$ gauge theories with an anti-symmetric hypermultiplet and $N_f$ fundmental hypermultiplets and $\theta=0$ for $1\le N_f\le 6$. Depending on the number of fundamental hypermultiplets, the number of instantons that we include to get a full answer varies. For low $N_f$ such as 1, 2, 3, we observed that including up to 2 instantons suffice for the final index to be a sum of characters of $E_{N_f+1}$ irreducible representations, but for high $N_f$ such as 4, 5, 6, we included up to 4 instantons for safety, and indeed the final index organizes itself to be a sum of characters of $E_{N_f+1}$. We have only analyzed up to $O(t^2)$ order. $N_f=2$ case is unsatisfactory, as we could not observe the exact decomposition of the index in terms of $E_{3}$ characters, but guessed the possible representations that are allowed given the information of the shifted $U(1)$ charges.  
\end{itemize}
\subsection{Adjoint ray operator index for G=Sp(1)}
Before presenting the result, let us briefly recall the global symmetry enhancement of 5d $\cN=1$ $G=Sp(1)$ gauge theory with $N_f$ fundamental hypermultiplets, as it is important to understand the overall structure of the index. The gauge theory has $SO(2N_f)$ flavor symmetry and $U(1)$ topological symmetry, which charges the instanton operators \cite{Lambert:2014jna,Tachikawa:2015mha}. The IR flavor symmetry group enhances to $E_{n}$ global symmetry in the UV, where $n=N_f+1$. Since we use the IR gauge theory description to compute the index, we need to track flavor fugacities $f_i$
and instanton fugacity $q$ to write down a sensible index in the UV. The precise map between the Cartan generators of the UV global symmetry and those of the IR global symmetry for each $N_f$ was presented in \cite{Kim:2012gu}. 
 
There is an additional subtlety associated to the instanton fugacity $q$ in the context of ray index. It was shown in \cite{Chang:2016iji} that on states that correspond to a fundamental ray operator, the U(1) instanton charge receives an anomalous contribution\footnote{See section 3.2 of \cite{Chang:2016iji} for detail.}
\ie
\frac{2}{8-N_f}.
\fe
In other words, the fundamental ray operator index has an overall factor of $q^{\frac{2}{8-N_f}}$. In the similar vein, we will observe the overall q-power shift for the adjoint ray index in the presence of $N_f$ fundamental hypers:
\ie\label{nfshift}\nonumber
2\times\frac{2}{8-N_f}.
\fe\\
{\bf{$N_f=0$ case:}}\\
\newline
Let us start with the case with no fundamental hypermultiplets$(N_f=0)$ in the IR gauge theory. Then, the result of \eqref{d4d4pexp} is
\ie\label{sp1ray}
\mathcal{I}^{\text{ray}}_{{Sp(1)}}=~&x^{-1}\left(1+\chi_3(q)t^2+(1+\chi_3(q))\chi_2(u)t^3+O(t^4)\right)\\
+&x^1\sqrt{q}\big(\chi_2(\sqrt{q})+(\chi_4(\sqrt{q})+\chi_2(\sqrt{q}))t^2+O(t^3)\big)\\
+&O(x^3) \,.
\fe
Note that $O(x^{-1})$ term of \eqref{sp1ray} successfully reproduces the superconformal index presented in section 4.1 of \cite{Hwang:2014uwa}, providing a consistency check for our computation. There is no $O(x^0)$ term as expected, since the theory does not contain any quark that can be attached to the end of a fundamental Wilson line. The $O(x^{1})$ term should be interpreted as index for the adjoint ray operators $\mathcal{I}^{\bf{adj}}_{\text{ray}}$. 
\ie
\cI^{\bf{adj}}_{\text{ray}}=\sqrt{q}\big(\chi_2(\sqrt{q})+(\chi_4(\sqrt{q})+\chi_2(\sqrt{q}))t^2+O(t^3)\big)
\fe
Notice that we do not find $\mathcal{I}_{\text{ray}}^{\bf{1}}$ contribution at $x^{1}$ order, as expected from the previous section. 

$N_f\geq1$ cases are essentially different from the $N_f=0$ case that we have just studied in the sense that there exist quarks on which the fundamental Wilson ray may end. Hence, we can test our OPE proposal for these cases.\\
\newline
{\bf{Adjoint ray operator index for 5d $E_2=SU(2)\times U(1)$ SCFT }}\\
\newline
For simplicity, we will only write down the adjoint ray index and refer to \cite{Chang:2016iji} for the information of the fundamental ray index.
\ie\label{sp1e2adjointray}
\mathcal{I}^{\text{ray}}=~&O(x^{-1})+O(x^0)\\
+&x\bigg(z^{\frac{1}{7}}\big(\chi_2(y)t^0+(\chi_4(y)+2\chi_2(y))t^2+\chi_2(u)(\chi_4(y)+3\chi_2(y))t^3+O(t^4)\big)\bigg)\\
+&O(x^2)
\fe
Here y and z are fugacities for UV flavor symmetry groups $SU(2)$ and $U(1)$ in $E_2$. They are related to the IR flavor symmetry fugacities $q$ and $y_1$ for $U(1)_I$ and $SO(2)$ as
\ie
y^2=qy_1,\quad z^2=\frac{y_1^7}{q}.
\fe

$E_2=SU(2)\times U(1)$ representations that appear in the adjoint ray index of $E_2$ theory are
\ie
{\bf{2_{1/7}}},\quad{\bf{4_{1/7}}}
\fe
where the subscripts are $U(1)_z$ charges. They are always 1 mod 7, and it matches with twice the $U(1)$ charge of fundamental ray operators, which is 4 mod 7.
\ie
4+4\text{ mod }7 = 1\text{ mod }7
\fe
This result is consistent with our proposal.\\
\newline
{\bf{Adjoint ray operator index for 5d $E_3=SU(3)\times SU(2)$ SCFT }}\\
\newline
\ie\label{sp1e3adjointray}
\mathcal{I}^{\text{ray}}=~&O(x^{-1})+O(x^0)\\
+&xq^{\frac{2}{3}}\bigg(\chi^{E_3}_{[1,0,0]}+\bigg[a_1\chi^{E_3}_{[4,0,0]}+a_2\chi^{E_3}_{[1,3,0]}\bigg]t^2+O(t^3)\bigg)\\
+&O(x^2)
\fe
where
\ie
\chi^{E_3}_{[1,0,0]}&=\frac{1}{q^{\frac{2}{3}}}+2q^{\frac{1}{3}},\\
\chi^{E_3}_{[4,0,0]}&=\frac{2}{q^{\frac{5}{3}}}+\frac{4}{q^{\frac{2}{3}}}+6q^{\frac{1}{3}}+3q^{\frac{4}{3}},\\
\chi^{E_3}_{[1,3,0]}&=\frac{4}{q^{\frac{5}{3}}}+\frac{8}{q^{\frac{2}{3}}}+6q^{\frac{1}{3}}+4q^{\frac{4}{3}}+2q^{\frac{7}{3}}.
\fe
Overall factor $q^{\frac{2}{3}}$ indicates the instanton charge shift\footnote{See \cite{Chang:2016iji} for more detail.} in the adjoint ray index:
\ie
\text{q-power shift for $N_f=2$ Adjoint ray index: }2\times\frac{2}{8-N_f}=\frac{2}{3}.
\fe
The {\bf{possible}}\footnote{$U(1)_I$ charges in the adjoint ray index after shifting with $q^{2/3}$, the list of possible $U(1)$ charges are $\frac{7-3k}{3}$, where $k\in\bZ^+$. Rescaling it by 3, we have $7-3k$. We select the proper representations of $E_3$ by looking at their $SO(4)\times U(1)$ tensor decomposition, and in particular the $U(1)$ charge of the decomposed representations. The {\bf{possible}} representations are those with U(1) charge $7-3k$ that appear in the adjoint ray index.} $E_3=SU(3)\times SU(2)$ representations that appear in the adjoint ray index of $E_3$ theory are 
\ie\label{e3}
{\bf{3}}=[1,0,0],\quad{\bf{15'}}=[4,0,0],\quad{\bf{24}}=[1,3,0]
\fe
Note that our notation for $E_3$ characters $[a,b,c]$ is equivalent to $[a,b]\times[c]$ of $SU(3)\times SU(2)$. $\bZ_3\times\bZ_2$ center of $SU(3)\times SU(2)$ can be determined by $a+b$ mod 3 and $c$ mod 2. In \eqref{e3}, we notice $c$ is always 0 mod 2, and $a+b$ mod $3$ is always 1. These central elements match with twice of $\mathbb{Z}_3\subset SU(3)$ and $\mathbb{Z}_2\subset SU(2)$ center charges of the fundamental ray operators, which are 2 mod 3 and 1 mod 2, respectively.
\ie
2+2\text{ mod }3~=~1\text{ mod }3,\quad 1+1\text{ mod }2~=~0\text{ mod }2
\fe
This is again consistent with our proposal in the previous subsection.\\
\newline
{\bf{Adjoint ray operator index for 5d $E_4=SU(5)$ SCFT }}\\
\newline
\ie\label{sp1e4adjointray}
\mathcal{I}^{\text{ray}}=~&O(x^{-1})+O(x^0)\\
+&xq^{\frac{4}{5}}\bigg(\chi^{E_4}_{[1,0,0,0]}+\bigg[\chi^{E_4}_{[2,0,0,1]}+\chi^{E_4}_{[0,1,0,1]}+\chi^{E_4}_{[1,0,0,0]}\bigg]t^2+O(t^3)\bigg)\\
+&O(x^2)
\fe
where
\ie
\chi^{E_4}_{[1,0,0,0]}&=\frac{1}{q^{\frac{4}{5}}}+4q^{\frac{1}{5}},\\
\chi^{E_4}_{[2,0,0,1]}&=\frac{4}{q^{\frac{9}{5}}}+\frac{16}{q^{\frac{4}{5}}}+40q^{\frac{1}{5}}+10q^{\frac{6}{5}},\\
\chi^{E_4}_{[0,1,0,1]}&=\frac{15}{q^{\frac{4}{5}}}+24q^{\frac{1}{5}}+6q^{\frac{6}{5}}.
\fe
Overall factor $q^{\frac{4}{5}}$ indicates the instanton charge shift in the adjoint ray index:
\ie
\text{q-power shift for $N_f=3$ Adjoint ray index: }2\times\frac{2}{8-N_f}=\frac{4}{5}.
\fe
The $E_4=SU(5)$ reps that appear in the adjoint ray index: 
\ie
{\bf{5}}=[1,0,0,0],\quad{\bf{45}}=[0,1,0,1],\quad{\bf{70}}=[2,0,0,1]
\fe
The representation $[a,b,c,d]$ corresponds to a Young diagram with total number of boxes $4a+3b+2c+d$. All the representations appear above has 4 mod 5 boxes. To see the $\bZ_3$ charges, we need to decompose each representation in terms of $SO(6)\times U(1)$ representations using the branching rule $SU(5)\rightarrow SO(6)\times U(1)$:
\ie
{\bf{5}}&={\bf{1}}_{-4}\oplus{\bf{4}}_{1}\\
{\bf{45}}&={\bf{15}}_{-4}\oplus{\bf{4}}_{1}\oplus{\bf{20}}_{1}\oplus{\bf{6}}_{6}\\
{\bf{70}}&={\bf{\bar4}}_{-9}\oplus{\bf{1}}_{-4}\oplus{\bf{15}}_{-4}\oplus{\bf{4}}_{1}\oplus{\bf{36}}_{1}\oplus{\bf{10}}_{6}
\fe
Reading $U(1)$ charges, we see they are all 1 mod 5. Therefore, they are charged 1 under $\bZ_5$. These central elements match with twice of $\mathbb{Z}_5$ center charge, 3 mod 5, of the fundamental ray operators.
\ie
3+3\text{ mod }5~=~1\text{ mod }5
\fe
This is consistent with our proposal in the previous subsection.\\
\newline
{\bf{Adjoint ray operator index for 5d $E_5=SO(10)$ SCFT }}\\
\newline
\ie\label{sp1e5adjointray}
\mathcal{I}^{\text{ray}}=~&O(x^{-1})+O(x^0)\\
+&xq^1\bigg(\chi^{E_5}_{[1,0,0,0,0]}+\bigg[8\chi^{E_5}_{[2,0,0,0,0]}+\chi^{E_5}_{[0,0,1,0,0]}+\chi^{E_5}_{[1,0,0,0,0]}-112\bigg]t^2+O(t^3)\bigg)\\
+&O(x^2)
\fe
where
\ie
\chi^{E_5}_{[1,0,0,0,0]}&=\frac{1}{q}+8+q,\\
\chi^{E_5}_{[2,0,0,0,0]}&=\frac{1}{q^2}+\frac{8}{q}+36+8q+q^2,\\
\chi^{E_5}_{[0,0,1,0,0]}&=\frac{28}{q}+64+28q
\fe
The overall factor $q^1$ indicates the instanton charge shift in the adjoint ray index:
\ie\label{nf4shift}
\text{q-power shift for $N_f=4$ Adjoint ray index: }2\times\frac{2}{8-N_f}=1.
\fe

The nontrivial $E_5=SO(10)$ representations that appear in the adjoint ray index are
\ie\label{so10}
{\bf{10}=[1,0,0,0,0]},\quad{\bf{54=[2,0,0,0,0]}},\quad{\bf{120=[0,0,1,0,0]}}.
\fe
$\bZ_4$(of Spin(10)) center charge can be computed by the shifted $U(1)$ charge plus 2 times $\bZ''_2$ charge, where $\bZ''_2$ is a subgroup of the center $\bZ'_2\times\bZ''_2$ of $Spin(8)$. As we have an integral q-charge shift in the adjoint index shown in \eqref{nf4shift}, the index is still an integral power series in q even after the shift, which indicates the shifted $U(1)$ charge is even. In the branching rule of $Spin(10)$ representations into representations of $Spin(8)\times U(1)$, even U(1) charge is paired with representations of Spin(8) that appear in tensor products of ${\bf{8}_c}$. The $\bZ''_2$ center charge of basic Spin(8) representations ${\bf{8}_v}$, ${\bf{8}_s}$, ${\bf{8}_c}$ are 0,1,1.

Looking at the decomposition\footnote{Note that ${\bf{8}_c}$, not the standard ${\bf{8}_v}$, appears in the above decompositions, since the embedding $Spin(8)\times U(1)$ is not the standard one, but related to it by triality. I thank Lakshya Bhardwaj who pointed out this subtlety.} of each Spin(10) representations in \eqref{so10},
\ie
{\bf{1}}&={\bf{1}},\\
{\bf{10}}&={\bf{1}}_2\oplus{\bf{1}}_{-2}\oplus{\bf{8}_c}_0\\
{\bf{54}}&={\bf{1}}_{-4}\oplus{\bf{8}_c}_{-2}\oplus{\bf{1}}_{0}\oplus{\bf{35}}_{0}\oplus{\bf{8}_c}_{-2}\oplus{\bf{1}}_{4}\\
{\bf{120}}&={\bf{28}}_{-2}\oplus{\bf{56}}_0\oplus{\bf{8}_c}_0\oplus{\bf{28}}_2,
\fe
we can read center charges: {\bf{1}}, {\bf{10}}, {\bf{54}} has 0, 2, 0, 2 $\bZ_4$ center charge. 

Out of two $\bZ_4$ center charges $\{0,2\}$ only $2$ can be matched with twice of $\mathbb{Z}_4$ center charge 3 mod 4, of fundamental ray operators.
\ie
3+3\text{ mod }4~=~2\text{ mod }4
\fe
The other adjoint ray operators charged as $0$ under the $\bZ_4$ center are expected to originate from two non-BPS operators in the sector of fundamental ray operators\footnote{I thank Lakshya Bhardwaj for pointing out this.}. With this understood, the result is consistent with our proposal in the previous subsection.
\\
\newline
{\bf{Adjoint ray operator index for 5d $E_6$ SCFT }}\\
\newline
\ie\label{sp1e6adjointray}
\mathcal{I}^{\text{ray}}=~&O(x^{-1})+O(x^0)\\
+&xq^{\frac{4}{3}}\bigg(\chi^{E_6}_{[1,0,0,0,0,0]}+\bigg[\chi^{E_6}_{[0,0,0,1,0,0]}+\chi^{E_6}_{[1,0,0,0,0,1]}+\chi^{E_6}_{[1,0,0,0,0,0]}-144\bigg]t^2+O(t^3)\bigg)\\
+&O(x^2)
\fe
where
\ie
\chi^{E_6}_{[1,0,0,0,0,0]}&=\frac{1}{q^{\frac{4}{3}}}+\frac{16}{q^{\frac{1}{3}}}+10q^{\frac{2}{3}},\\
\chi^{E_6}_{[0,0,0,1,0,0]}&=\frac{45}{q^{\frac{4}{3}}}+\frac{160}{q^{\frac{1}{3}}}+130q^{\frac{2}{3}}+16q^{\frac{5}{3}},\\
\chi^{E_6}_{[1,0,0,0,0,1]}&=\frac{16}{q^{\frac{7}{3}}}+\frac{256}{q^{\frac{4}{3}}}+\frac{592}{q^{\frac{1}{3}}}+576q^{\frac{2}{3}}+144q^{\frac{5}{3}}.
\fe
Overall factor $q^{\frac{4}{3}}$ indicates the instanton charge shift in the adjoint ray index:
\ie\label{nf5shift}
\text{q-power shift for $N_f=5$ Adjoint ray index: }2\times\frac{2}{8-N_f}=\frac{4}{3}.
\fe

The nontrivial $E_6$ representations that appear in the adjoint ray index are
\ie
{\bf{27=[1,0,0,0,0,0]}},\quad{\bf{351=[0,0,0,1,0,0]}},\quad{\bf{1728=[1,0,0,0,0,1]}}
\fe
Once again, we need to look the decomposition to read off $\bZ_3$ center charge of each representation.
\ie
{\bf{27}}&={\bf{1}}_{-4}\oplus{\bf{10}}_{2}\oplus{\bf{16}}_{-1}\\
{\bf{351}}&={\bf{45}}_{-4}\oplus{\bf{16}}_{-1}\oplus{\bf{144}}_{-1}\oplus{\bf{10}}_{2}\oplus{\bf{120}}_{2}\oplus{\bf{\bar 16}}_{5}\\
{\bf{1728}}&={\bf{\bar 16}}_{-7}\oplus{\bf{1}}_{-4}\oplus{\bf{45}}_{-4}\oplus{\bf{210}}_{-4}\oplus2{\bf{\bar 16}}_{-1}\oplus{\bf{144}}_{-1}\oplus{\bf{560}}_{-1}\oplus{\bf{10}}_{2}\oplus{\bf{120}}_{2}\oplus{\bf{\bar126}}_{2}\oplus{\bf{144}}_{5}
\fe
Their $U(1)$ charges are all 2 mod 3, from which we read off $\bZ_3$ charge 2. These central elements match with twice of $\mathbb{Z}_3$ center charge, 1 mod 3, of the fundamental ray operators.
\ie
1+1\text{ mod }3~=~2\text{ mod }3.
\fe
Result is consistent with our proposal in the previous subsection.
 \\
\newline
{\bf{Adjoint ray operator index for 5d $E_7$ SCFT }}\\
\newline
\ie\label{sp1e7adjointray}
\mathcal{I}^{\text{ray}}=~&O(x^{-1})+O(x^0)+xq^2\bigg(\chi^{E_7}_{[1,0,0,0,0,0,0]}+\chi^{E_7}_{[1,0,0,0,0,0,0]}t+O(t^3)\bigg)+O(x^2)
\fe
where
\ie
\chi^{E_7}_{[1,0,0,0,0,0]}&=\frac{1}{q^{2}}+\frac{32}{q}+68+32q+q^2.
\fe
Overall factor $q^2$ indicates the instanton charge shift in the adjoint ray index:
\ie\label{nf6shift}
\text{q-power shift for $N_f=6$ Adjoint ray index: }2\times\frac{2}{8-N_f}=2.
\fe

The nontrivial $E_7$ representations that appear in the adjoint ray index are
\ie
{\bf{133=[1,0,0,0,0,0]}}
\fe
Note that {\bf{133}} is the representation that appears in the superconformal index of $E_7$ theory. Since we know in general the superconformal index only contains $E_n$ representations with trivial center charges, this implies that {\bf{133}} has 0 charge under $\bZ_2$ center. 

These central elements match with twice of $\mathbb{Z}_2$ center charge, 1 mod 2, of the fundamental ray operators.
\ie
1+1\text{ mod }2~=~0\text{ mod }2.
\fe
Result is consistent with our proposal in the previous subsection.
\subsection{Adjoint ray operator index for G=Sp(2), Sp(3)}
For $Sp(N)$ with $N\geq2$, we need to treat them separately. Consider, for instance, $N=2$. The $O(x^{-N+2})$ terms encode ray operators that transform under $Sp(2)_{D4}\times Sp(1)_{D4'}$ as
\ie
(\bf{4},\bf{2})\wedge(\bf{4},\bf{2})=(\bf{Adj},\bf{1})\oplus(\bf{Asym},\bf{3})\oplus(\bf{1},\bf{3}) \,.
\fe
Extra states not related to D4-brane, $(\bf{1},\bf{3})$, are decoupled from the index as before, so we are left with
\ie
(\bf{Adj},\bf{1})\oplus(\bf{Asym},\bf{3})\,.
\fe
Therefore,
\ie\label{genseries}
\cI^{\text{ray}}_{Sp(N\geq2)}=x^{-N}\left(\cI_{SCI}+x\cI^{\bf{fund}}_{\text{ray}}+x^2(\cI^{\bf{adj}}_{\text{ray}}+\cI^{\bf{asym}}_{\text{ray}})+\ldots\right)\,.
\fe
In this subsection, we will only present the result for $N_f=0$ case.\\
\newline
{\bf{Adjoint ray operator index for G=Sp(2)}}\\
\newline
\ie\label{sp2ray}
\mathcal{I}^{\text{ray}}_{{Sp(2)}}=~&x^{-2}\big(1+\chi_2(v)t+(\chi_2(u)\chi_2(v)+2\chi_3(v)+\chi_3(q))t^2+O(t^3)\big)\\
+&x^0\bigg(\sqrt{q}\chi_2(\sqrt{q})+(\sqrt{q}\chi_2(v)\chi_2(\sqrt{q})+{\chi_2(v)}+2\sqrt{q}\chi_2(\sqrt{q})-2)t\\
+&\big(\sqrt{q}(\chi_4(\sqrt{q})+\chi_2(v)\chi_2(\sqrt{q})(2\chi_2(u)+2\chi_3(v)))+{\chi_2(v)^2-2\chi_2(v)}\big)t^2\bigg)\\
+&O(x^2)
\fe
Note that $O(x^{-2})$ term in \eqref{sp2ray} successfully reproduces the superconformal index presented in section 4.2 of \cite{Hwang:2014uwa}, providing a consistency check for our computation. The $O(x^{0})$ term contains contributions both from $\mathcal{I}^{\bf{adj}}_{\text{ray}}$ and $\cI^{\bf{asym}}_{\text{ray}}$.  \\
\newline
{\bf{Adjoint ray operator index for G=Sp(3)}}\\
\newline
\ie
\mathcal{I}^{\text{ray}}_{{Sp(3)}}=~&x^{-2}\big(1+\chi_2(v)t+(\chi_2(u)\chi_2(v)+2\chi_3(v)+\chi_3(q))t^2+O(t^3)\big)\\
+&x^0\bigg(\sqrt{q}\chi_2(\sqrt{q})+(\sqrt{q}\chi_2(v)\chi_2(\sqrt{q})+{\chi_2(v)}+2\sqrt{q}\chi_2(\sqrt{q})-2)t\\
+&\big(\sqrt{q}(3\chi_2(\sqrt{q})+\chi_4(\sqrt{q})+\chi_2(v)\chi_2(\sqrt{q})(2\chi_2(u)+2\chi_3(v)))-4+{2\chi_2(v)^2-2\chi_2(v)}\big)t^2\bigg)\\
+&O(x^2)
\fe
Similar to $G=Sp(2)$ case, $O(x^{0})$ term contains contributions both from $\mathcal{I}^{\bf{adj}}_{\text{ray}}$ and $\cI^{\bf{asym}}_{ray}$. 

As a final remark, it would be interesting to see if the index computation given in this work can be reproduced in a geometric way using the realization of the 5d SCFT in terms of M-theory on Calabi-Yau 3-fold \cite{Douglas:1996xp,Intriligator:1997pq}(for instance, see \cite{Bhardwaj:2020ruf,Bhardwaj:2020avz}.) Moreover, it would be nice to make contact with recent works on the global form of flavor symmetries and 2-group symmetries in the line of \cite{BenettiGenolini:2020doj,Apruzzi:2021vcu}. Lastly, it is curious if there exist universal expressions for the ray index; this turns out to be the case for the superconformal index \cite{Song:2021dhu}. \section*{Acknowledgments}
I thank Fabio Apruzzi, Lakshya Bhardwaj, Hee-Cheol Kim, Joonho Kim, Sakura Schäfer-Nameki for discussion and useful email correspondence. I am especially grateful to Lakshya Bhardwaj for asking important questions and for many insightful comments on the draft. I am supported by ERC Grants 682608 and 864828.

\providecommand{\href}[2]{#2}\begingroup\raggedright
    
    \endgroup
\end{document}